\documentclass[3p,times,procedia]{elsarticle}
\usepackage{nupha_ecrc}

\volume{00}

\firstpage{1}

\journalname{Nuclear Physics A}

\runauth{C. Gale et al.}

\jid{nupha}

\jnltitlelogo{Nuclear Physics A}

\usepackage{amssymb}

\usepackage[figuresright]{rotating}

\begin{document}

\begin{frontmatter}



\dochead{XXVIIIth International Conference on Ultrarelativistic Nucleus-Nucleus Collisions\\ (Quark Matter 2019)}

\title{Probing Early-Time Dynamics and Quark-Gluon Plasma Transport Properties with Photons and Hadrons}


\author[McGill]{Charles Gale}
\author[Duke]{Jean-Fran\c{c}ois Paquet}
\author[BNL]{Bj\"orn Schenke}
\author[WSU]{Chun Shen}
\address[McGill]{Department of Physics, 3600 University Street, Montreal, QC, H3A 2T8, Canada}
\address[Duke]{Department of Physics, Duke University, Durham, NC 27708, USA}
\address[BNL] {Physics Department, Brookhaven National Laboratory, Upton, NY 11973, USA}
\address[WSU]{Department of Physics and Astronomy, Wayne State University, Detroit, MI 48201, USA}

\begin{abstract}
We use  a hybrid  model which relies on   QCD effective kinetic theory -- K{\o}MP{\o}ST -- to dynamically bridge the gap between IP-Glasma initial states and viscous hydrodynamics, for the theoretical interpretation of heavy-ion collision results. The hydrodynamic phase is then followed by dynamical freeze-out handled by UrQMD. It is observed that the new pre-equilibrium/pre-hydro phase does influence the extraction of transport coefficients. 
We also show how the early dynamics is reflected in photon observables. 
\end{abstract}


\end{frontmatter}


\section{Introduction}
In addition to the observation at RHIC and at the LHC of the exotic state of matter known as the Quark-Gluon Plasma (QGP), one of the remarkable developments in theory has been the success in modelling this medium  with relativistic viscous hydrodynamics \cite{Gale:2013da}. More specifically, the collectivity of the QGP is typically analyzed by performing a Fourier expansion in azimuthal angle of the differential cross-section of measured species. The expansion coefficients are known as flow parameters, and their value can be extracted by experiments and calculated by theory \cite{Heinz:2013th}. However, it is clear from a theoretical point of view that the time-evolution of the colliding system can not be entirely ascribed to fluid dynamics, and that the space-time history of a collision event will need to be separated in distinct eras. At high colliding energies, the very early stages  of heavy-ion collisions are represented by a gluon-dominated parton population. This high-occupancy medium is classical in nature and its evolution obeys the classical Yang-Mills equations: this defines the IP-Glasma \cite{Schenke:2012wb}. The majority of phenomenological applications so far have used a version of this early-time evolution of boost-invariant gluon fields coupled to hydrodynamics at a switching time of $\tau \sim 1/Q_s$, when $Q_s$ is the saturation scale: the energy scale where the non-linearity of QCD sets in and will manifest itself in the gluon population \cite{Gale:2012rq}. The modern hydrodynamics evolution now takes into account dissipative processes and is followed by hadronization, which then initiates a hadronic transport. This assembly of theoretical tools  defines  a class of {\it hybrid models}, which represent the state-of-the-art used to analyze the phenomenology of relativistic heavy-ion collisions. Recently, a coarse-grained QCD effective kinetic theory approach -- K{\o}MP{\o}ST -- has been designed to interpolate between IP-Glasma and fluid dynamics \cite{Kurkela:2018vqr}. It is the purpose of this work to build on this development and to define a new hybrid model which strings together IP-Glasma-K{\o}MP{\o}ST-MUSIC-UrQMD and to assess the influence of the pre-hydro phase on hadronic and electromagnetic observables.

\section{The model and results}
Our hybrid approach can be summarized as follows: an initial IP-Glasma followed by K{\o}MP{\o}ST at a switching proper time $\tau$~=~0.1 fm/c.  MUSIC, a second-order fluid-dynamical simulation\footnote{http://www.physics.mcgill.ca/music/},  takes over at $\tau$ = 0.8 fm/c. Hadronization followed by UrQMD occurs when the local energy density reaches 0.18 GeV/fm$^3$. The specific value of the switching times will be explored in more detail in future work, but  current values are suggested by Ref.~\cite{Kurkela:2018vqr}, and are consistent with those  in many phenomenological applications.\footnote{For the comparison calculations shown in Fig.~\ref{hadrons} which do not contain K{\o}MP{\o}ST, the switching time between IP-Glasma and MUSIC was 0.4 fm/c \cite{Gale:2012rq}.} Importantly,  all elements of $T^{\mu \nu}$, the energy-momentum tensor, are handed over from one stage to the next, including  shear and bulk viscous components. For this first phenomenological application of K{\o}MP{\o}ST, we adopt a constant value for the specific shear viscosity, $\eta/s$ and a temperature-dependent $\zeta/s$, see next subsection.

\subsection{Hadrons}
\begin{figure}[th!]
\begin{minipage}{0.70\textwidth}
\begin{center}
\includegraphics[width=0.48\textwidth]{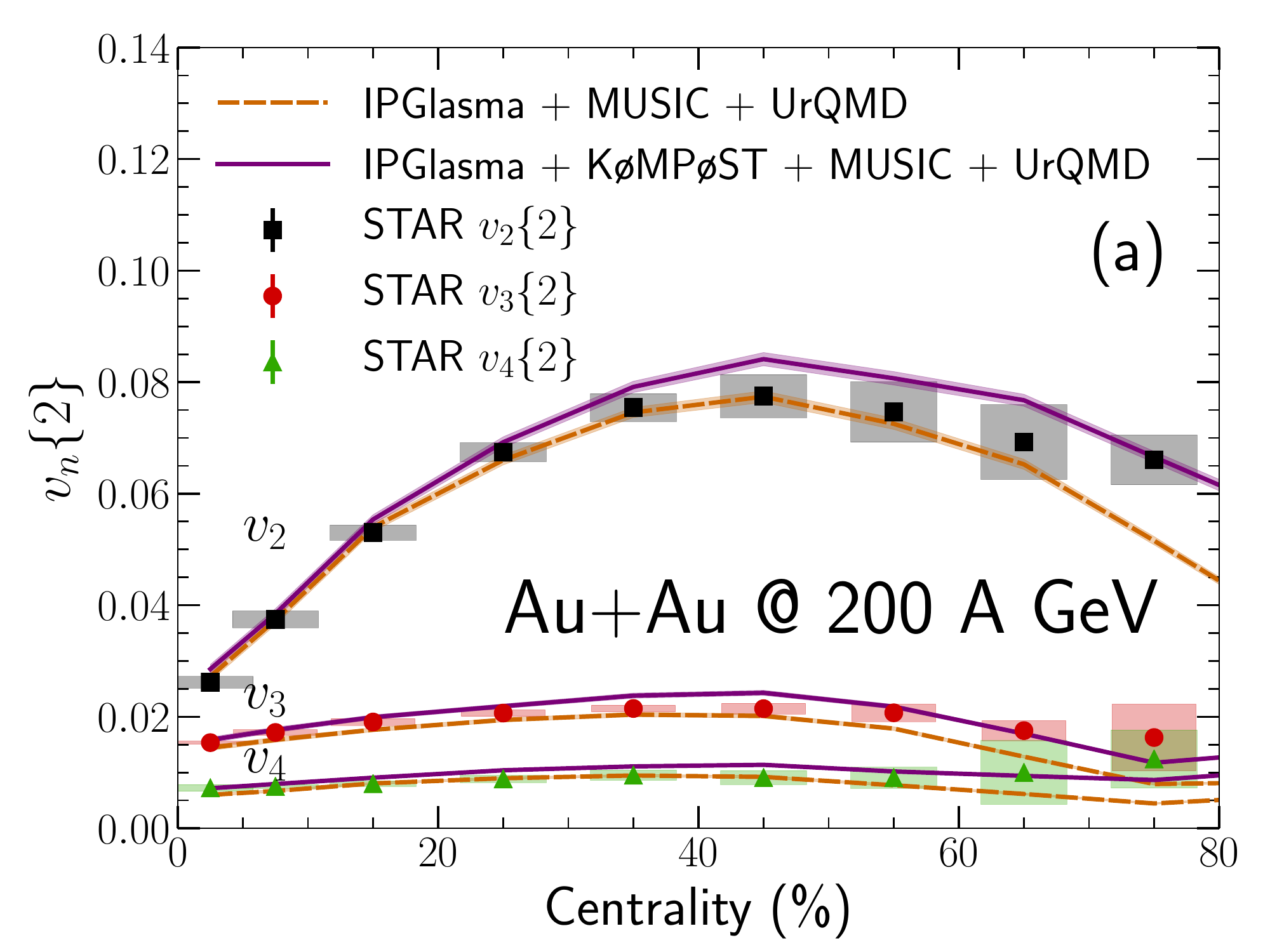}
\hspace*{0.051cm}\includegraphics[width=0.48\textwidth]{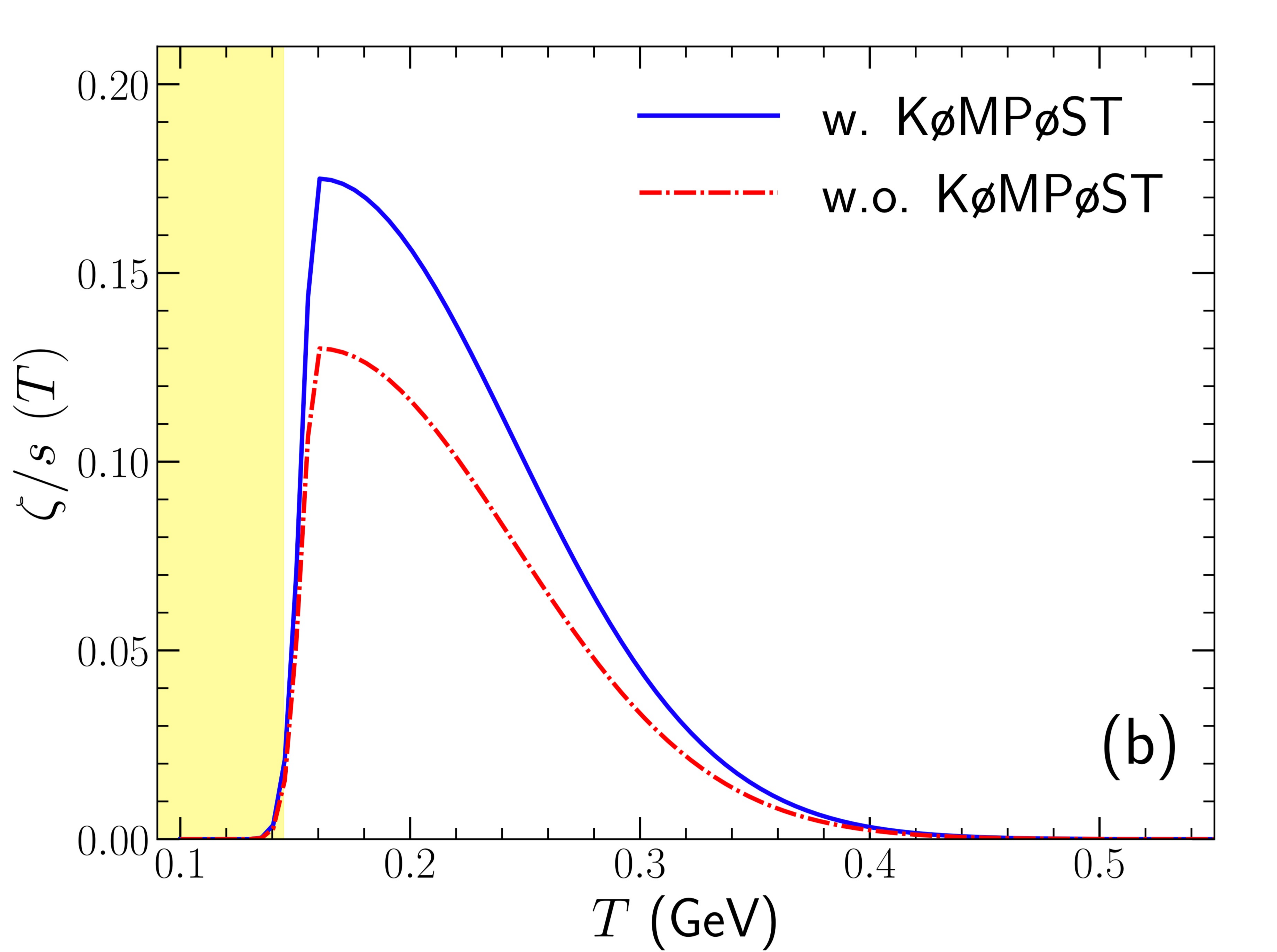}\\
\includegraphics[width=0.48\textwidth]{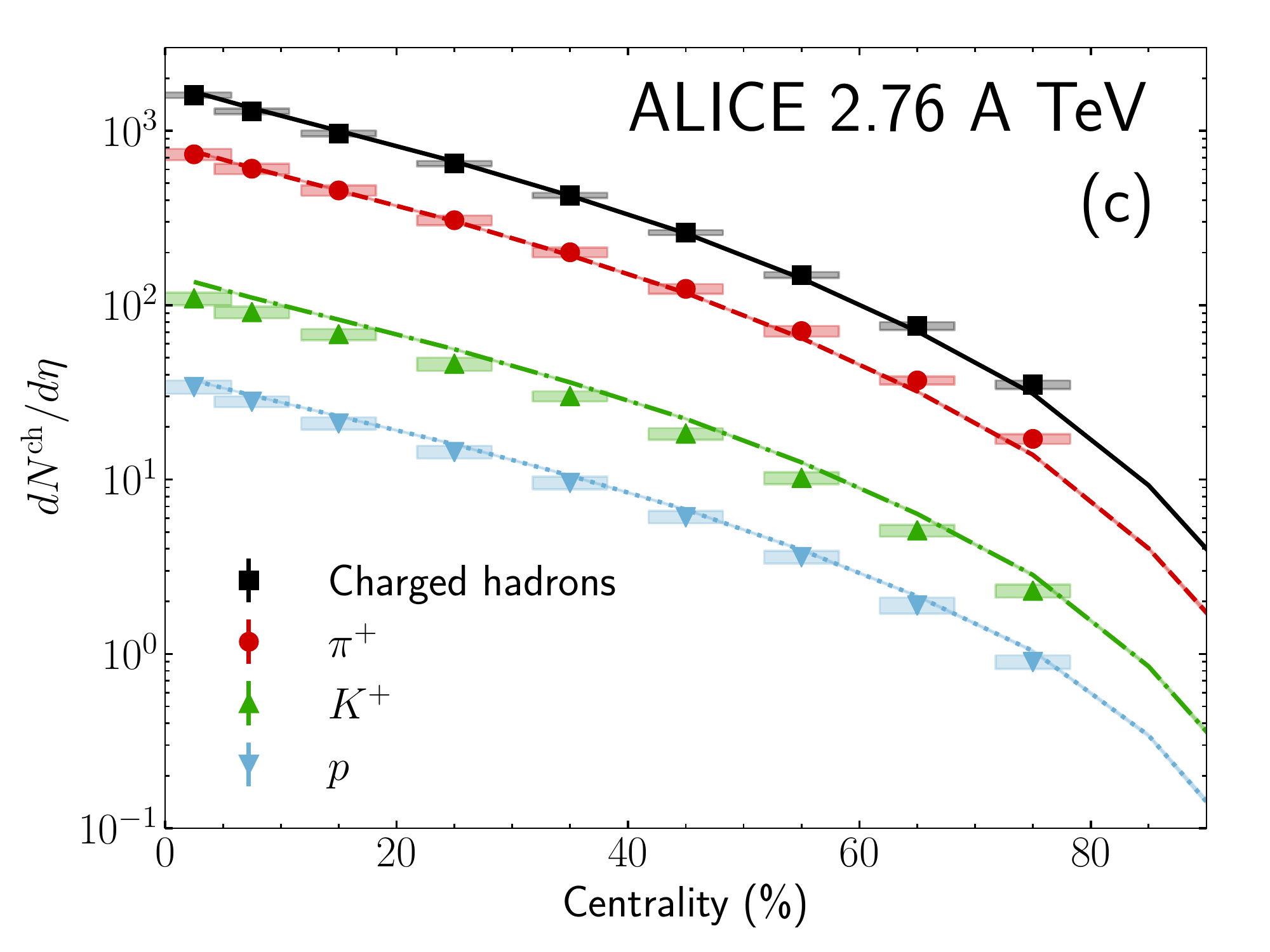}
\includegraphics[width=0.48\textwidth]{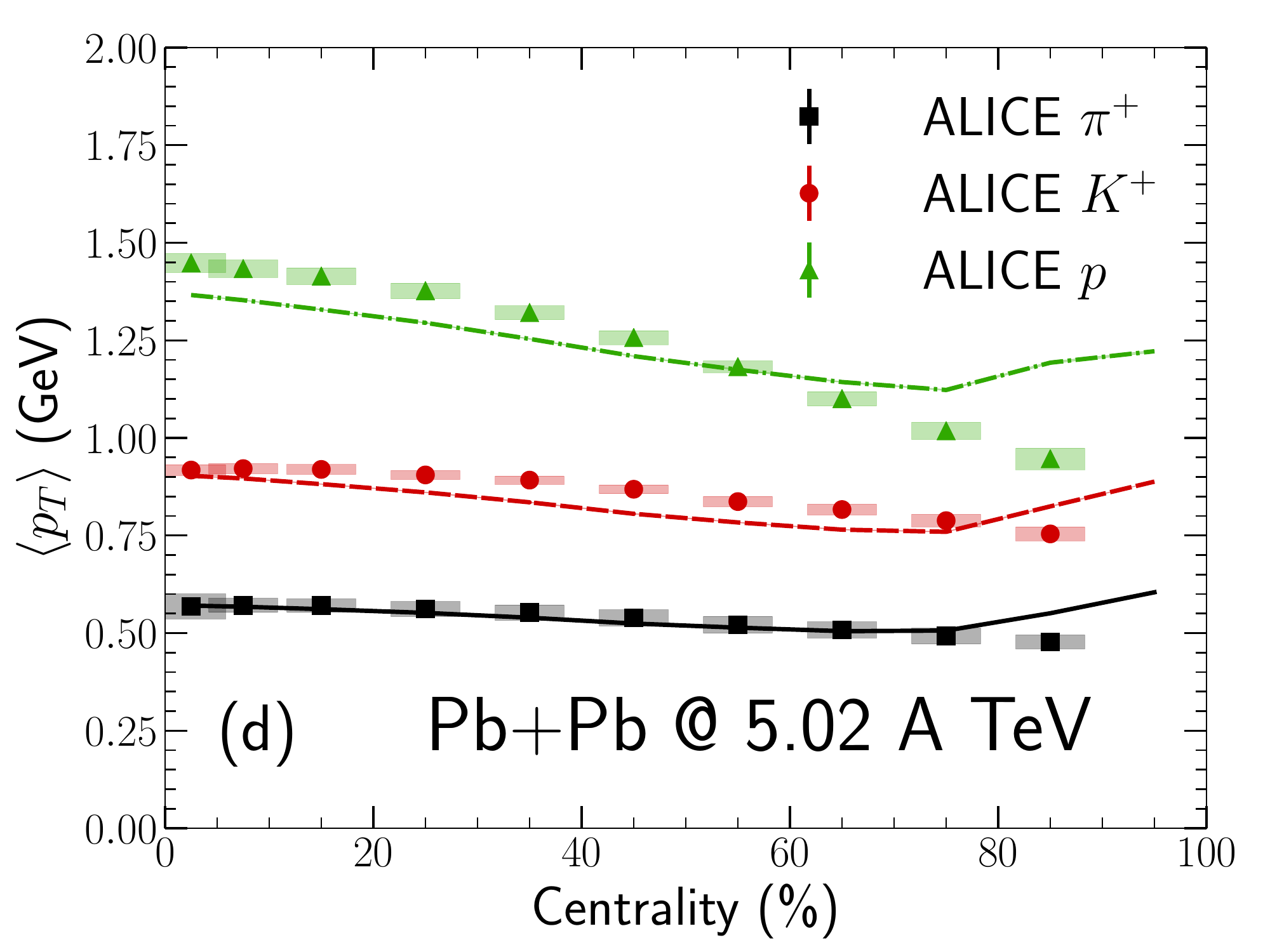} 
\end{center}
\end{minipage}
\begin{minipage}{0.29\textwidth}
\caption{A collection of hadronic results obtained with our hybrid approach compared to data. (a) Flow observables  as a function of collision centrality as measured by the STAR Collaboration at RHIC \cite{Adamczyk:2017hdl}, together with calculations with the hybrid model described herein and with an approach where the effective kinetic theory phase is absent \cite{Schenke:2020unx}. (b) The bulk viscosity per entropy density favoured by the data, depending on whether K{\o}MP{\o}ST is there or not. The region in yellow represents the phase ascribed to UrQMD in this work. (c) Multiplicity distribution of charged hadrons for Pb + Pb at 2.76 TeV$/A$.  (d) Average transverse momentum of identified hadrons for Pb + Pb at 5.02 TeV$/A$. Data are from the ALICE collaboration \cite{Abelev:2013vea, Acharya:2019yoi}.}
\label{hadrons}
\end{minipage}
\end{figure}

A first important result of our study is that one observes an effect on the bulk viscosity deduced from data analyses, depending on whether or not there is a K{\o}MP{\o}ST phase: see the top row of Fig.~\ref{hadrons}. The pre-hydro evolution will therefore influence the extraction  of transport coefficients of  strongly interacting matter. This is seen in Fig.~\ref{hadrons}a. For each of those cases, the specific bulk viscosity requested by the hadronic data for the fluid dynamical phase is plotted in Fig.~\ref{hadrons}b. 

The details of the calculation will appear separately \cite{Galeetal}, but from Fig.~\ref{hadrons}, it appears clear that our hybrid model which incorporates a non-equilibrium kinetic theory evolution can yield appropriate hadronic phenomenology for multiple collision systems over a range of energy that spans an order of magnitude, without readjusting shear and bulk viscosity. 

\subsection{Photons}
\begin{figure}[ht!]
\begin{minipage}{0.70\textwidth}
\begin{center}
    \includegraphics[width=0.48\textwidth]{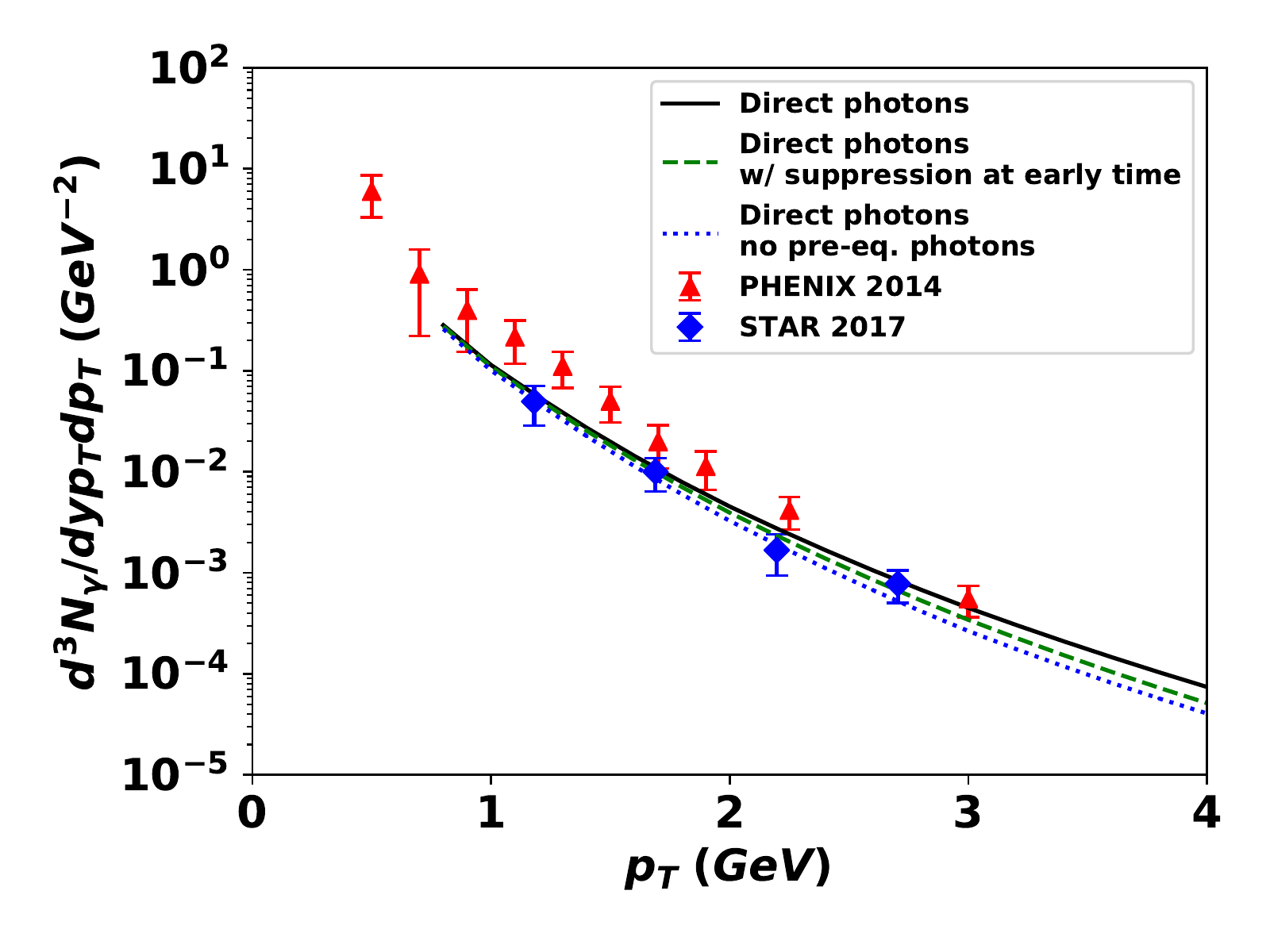}  \llap{
    	\parbox[b]{1.6in}{(a)\\\rule{0ex}{0.3in}
    }}
    \includegraphics[width=0.48\textwidth]{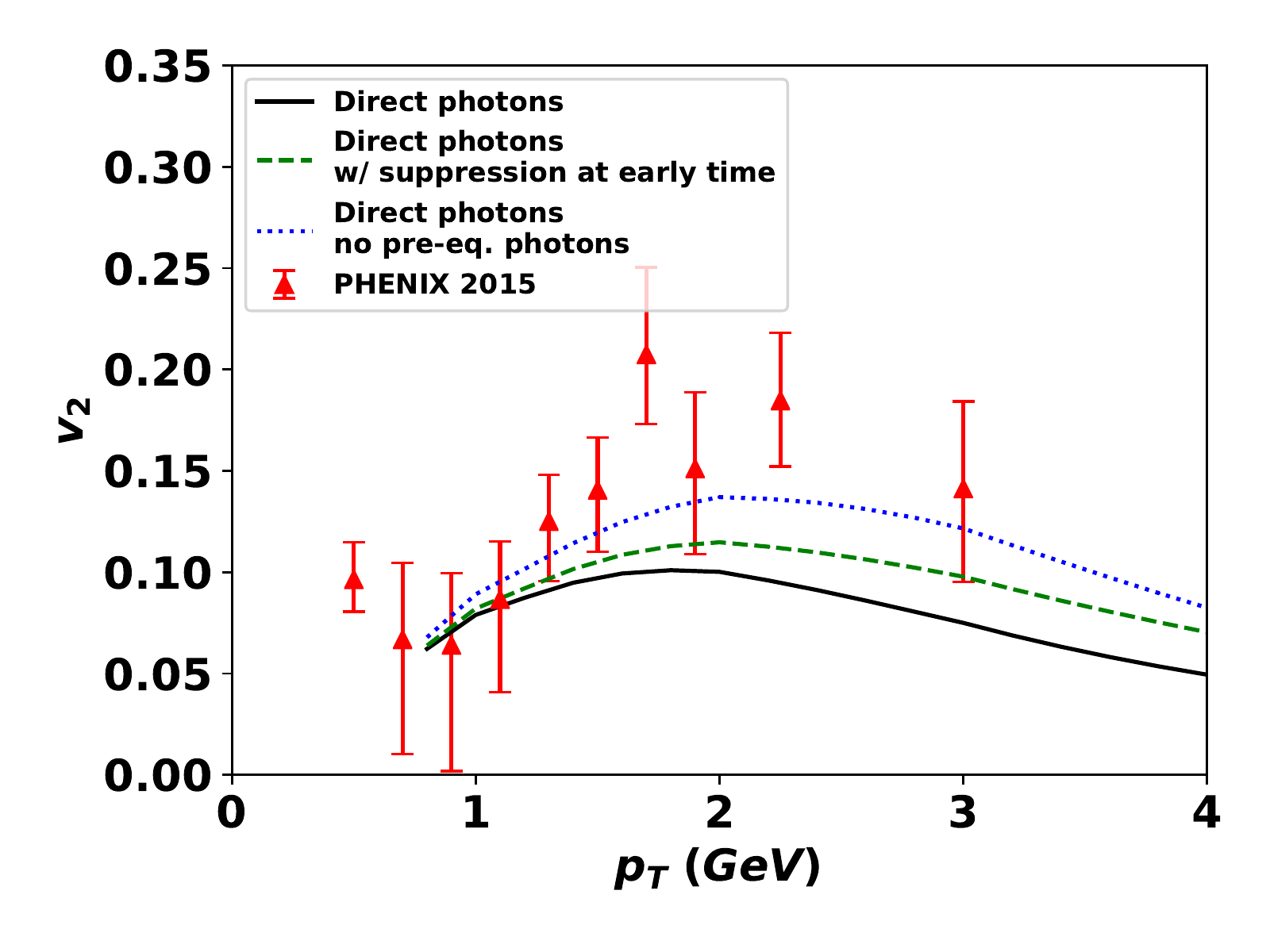} \llap{
    	\parbox[b]{1.62in}{(b)\\\rule{0ex}{0.3in}
    }} \\
    \includegraphics[width=0.48\textwidth]{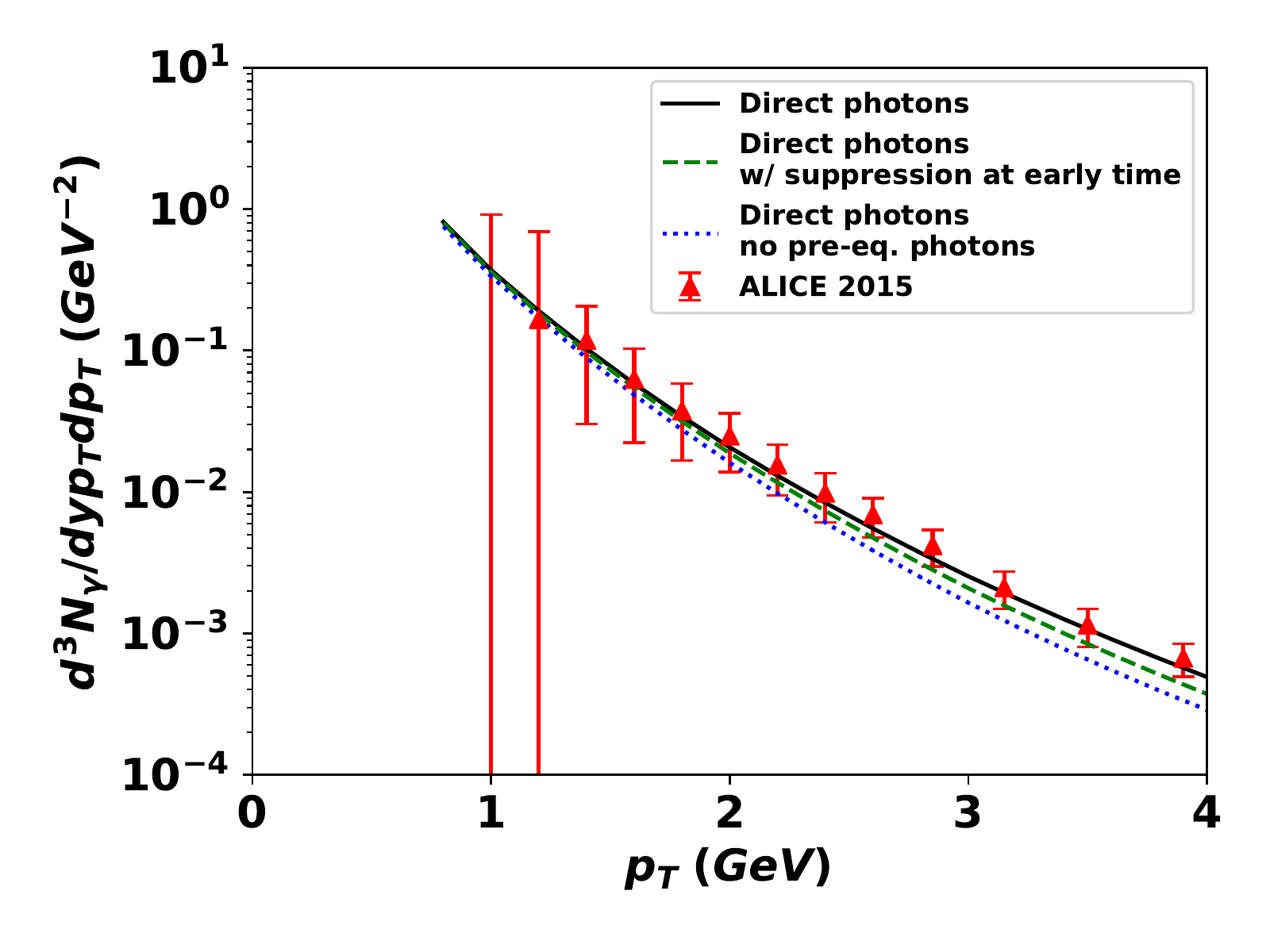} \llap{
    	\parbox[b]{1.6in}{(c)\\\rule{0ex}{0.3in}
    }}
    \includegraphics[width=0.48\textwidth]{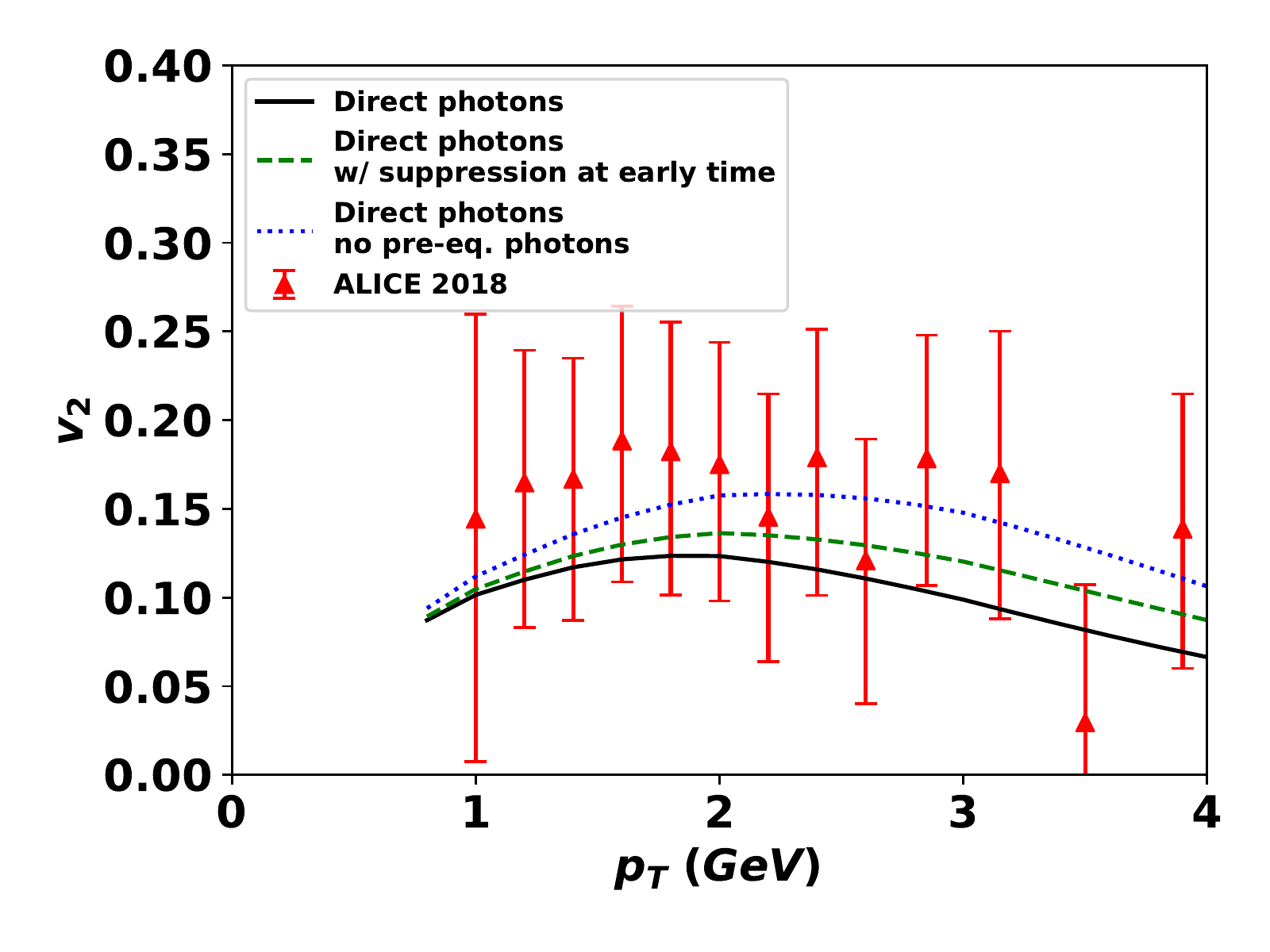}  \llap{
    	\parbox[b]{1.62in}{(d)\\\rule{0ex}{0.3in}
    }}
\end{center}
\end{minipage}
\begin{minipage}{0.28\textwidth}
\caption{This shows the spectrum (a) and $v_2$ (b) for direct photons in Au + Au collisions at $\sqrt{s}$ = 200 A GeV. The bottom row shows the same quantitiees (spectrum (c) and $v_2$ (d)) for Pb + Pb collisions at $\sqrt{s}$ = 2.76 A TeV. All plots here refer to a $20 - 40$\% centrality class. Data are from the PHENIX \cite{Adare:2014fwh,Adare:2015lcd} and STAR \cite{STAR:2016use} collaborations at RHIC, 
and from the ALICE \cite{Acharya:2018bdy,Adam:2015lda} collaboration at the LHC.}
\label{photon_fig}
\end{minipage}
\end{figure}

We now turn to a discussion on photon production. Owing to the relative size of the electromagnetic fine structure constant compared to its strong interaction counterpart, electromagnetic radiation will escape the medium upon creation and will reflect all epochs of the collision process \cite{Gale:2009gc,Tripolt:2020dac}. Traditionally, the vast majority of photon calculations account for the radiation from the very first nucleon-nucleon collisions and for that emitted throughout the fluid dynamical evolution  \cite{Paquet:2015lta}. Closer attention is now being paid to the late stages \cite{Schafer:2020vvw}, and some results  \cite{Churchill:2020yny,Greif:2016jeb} now also include the photons from early, pre-equilibrium, pre-hydro phases like the one considered here and studied using  K{\o}MP{\o}ST. The results shown in this work assume that, given $T^{\mu \nu}$ and a lattice-QCD equation of state, a local temperature can be extracted and used in calculated photon-producing rates.  In addition, non-viscous rates are used and the photons generated in the hadronic transport phase are simulated by evolving MUSIC down to a freeze-out temperature $T = 105$ MeV \cite{Paquet:2015lta}. Attention needs to be paid to partonic chemistry: recall that charged particles need to be present to produce photons, and  IP-Glasma and K{\o}MP{\o}ST are gluon-dominated in their current respective implementation. In reality, the fermion phase space will be populated dynamically via inelastic processes. We simulate this effect by implementing a smooth transition between a gluon-rich medium and a chemically equilibrated one, during the time interval when the effective kinetic theory is used \cite{Galeetal}. The results of these photon calculations are shown in Fig.~\ref{photon_fig}, which highlights  the photon spectrum and the photon elliptical flow.
 The dashed line shows the direct photon signal with ``early time suppression'': the correction attributed to the gluon-dominated beginning of the K{\o}MP{\o}ST phase. This is the  ``final'' photon result of this paper. The solid line is the direct photon signal without this correction. Finally, omitting completely the radiation from the effective kinetic theory period yields the  dotted line (note that this latter case has thermal photons emission starting at $\tau$ = 0.8 fm/c). All curves shown contain photons from primordial nucleon-nucleon collisions, scaled up by $N_{\rm coll}$, and calculated with pQCD at NLO with isospin and nuclear corrections \cite{Kovarik:2015cma}. The data shown are from the PHENIX and STAR collaborations at RHIC, and from the ALICE collaboration at the LHC. The presence of the ``pre-hydro photons'' is manifest, especially at higher $p_T$'s, where they typically represent $\sim10 - 20\%$ of the net spectrum. At RHIC, the calculated photon spectrum lies below the PHENIX data but is consistent with STAR measurements. The RHIC $v_2$ measurements from PHENIX lie systematically above the calculations, but are statistically consistent with the model results, except for two data points. The ALICE photon spectrum admits the presence of the K{\o}MP{\o}ST photons, and even has the potential to distinguish between cases with and without early chemical suppression. Both these scenarios fit the ALICE photon $v_2$, as they both  lie within the uncertainty bounds. A cautionary reminder:  photon spectra and $v_2$ evaluated in the theoretical hybrid approach outlined here do not yet account for viscous effects \cite{Paquet:2015lta}. These results should therefore be treated as representing preliminary estimates,  albeit  promising ones. 

\section{Conclusion}

The calculations reported here represent the first realistic integration of K{\o}MP{\o}ST in a modern hybrid approach, which consists of IP-Glasma, viscous MUSIC, and UrQMD. We have shown that hadronic and photon data are consistent with the presence of a pre-hydro phase as characterized by effective kinetic theory. More specifically, that phase has a direct influence on the extraction of QCD transport parameters -- notably the ratio of bulk viscosity over entropy density $\zeta/s$, and has an influence on the photon transverse momentum spectrum and flow coefficients. The degree to which  data unequivocally requires a pre-equilibrium epoch will eventually be quantified by state-of-the-art multivariable analyses like those recently discussed in Ref. \cite{Shen:2020gef}. However, {\it theory} currently demands this phase in order for IP-Glasma initial states to dynamically evolve into fluid-like systems. \\

\noindent {\bf Acknowledgements} This work was funded in part by the Natural Sciences and Engineering Research Council of Canada, by the U.S. Department of Energy (DOE) under grant numbers de-fg02-05ER41367, de-sc0012704, and de-sc0013460, and within the framework of the Beam Energy Scan Theory (BEST) Topical Collaboration.
 




\bibliographystyle{elsarticle-num}
\bibliography{QM_bib.bib}







\end{document}